\documentclass[pra, showkeys, twocolumn, floatfix, superscriptaddress]{revtex4}

\usepackage{graphicx}
\usepackage{amsmath, amsfonts, amssymb, bm}
\usepackage{nicefrac}

\usepackage[]{color}
\usepackage{amstext}
\usepackage{mathrsfs}
\usepackage[]{psfrag}

\usepackage[normalem]{ulem}

\newcommand{\dirac}{\displaystyle{\not}}
\newcommand{\e}{\text{e}}

\psfragscanoff
\providecommand{\abs}[1]{\lvert#1\rvert}

\begin{document}
\title{Higgs Boson Creation in Laser-Boosted Lepton Collisions}
\author{Sarah J. M\"uller}
\affiliation{Max-Planck-Institut f\"ur Kernphysik, Saupfercheckweg 1, D-69117 Heidelberg, Germany}
\author{Christoph H. Keitel}
\affiliation{Max-Planck-Institut f\"ur Kernphysik, Saupfercheckweg 1, D-69117 Heidelberg, Germany}
\author{Carsten M\"uller}
\affiliation{Max-Planck-Institut f\"ur Kernphysik, Saupfercheckweg 1, D-69117 Heidelberg, Germany}\affiliation{Institut f\"ur Theoretische Physik I, Heinrich-Heine-Universit\"at D\"usseldorf, Universit\"atsstr. 1, 40225 D\"usseldorf, Germany}

\begin{abstract}%
Electroweak processes in high-energy lepton collisions are
considered in a situation where the incident center-of-mass energy
lies below the reaction threshold, but is boosted to the required
level by subsequent laser acceleration. Within the framework of
laser-dressed quantum field theory, we study the laser-boosted
process {$\ell^+\ell^-\to HZ^0$} in detail and specify the technical
demands needed for its experimental realization. Further, we outline possible qualitative differences to field-free processes regarding the detection of the produced Higgs bosons. 
\end{abstract}


\maketitle

\begin{figure}[bt]
 \centering
\includegraphics[width=0.45\textwidth]{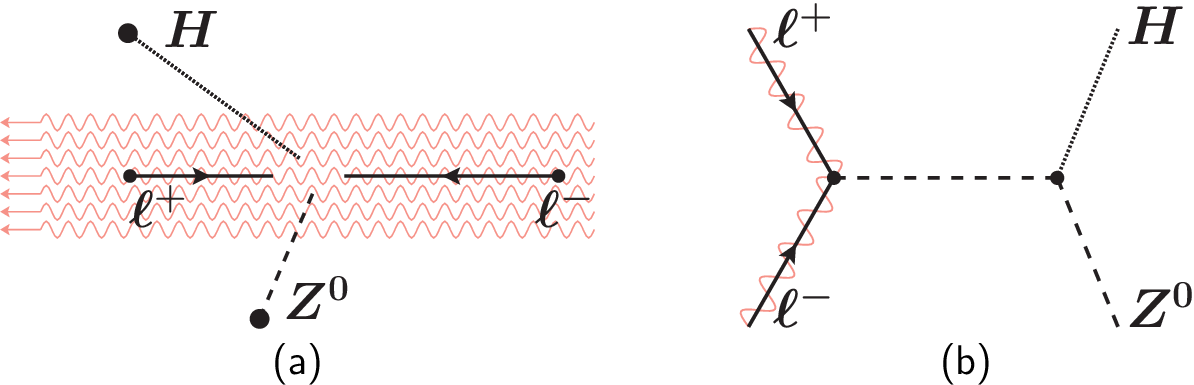}
\caption{(Color online) Schematic view of the considered process {$\ell^+\ell^-\to HZ^0$} inside a laser field (a) and its Feynman diagram (b). The wiggled lines indicate that the colliding leptons are described by laser-dressed states.}\label{Scheme}
\end{figure}

Search for the Higgs boson is one of the primary tasks of the Large
Hadron Collider (LHC) at CERN. It relies on collisions of few-TeV proton
beams produced in a 27 km long ring accelerator. The Higgs boson has been
predicted as physical manifestation of the Higgs mechanism which endues
the $W$ and $Z$ bosons with their mass by way of electroweak symmetry
breaking. The Higgs boson recently discovered at the LHC \cite{Higgs} shows standard-model like properties in all aspects that are accessible by the LHC experiment \cite{Higgs2}. However, further investigations of its couplings to other particles and to itself are necessary in order to establish whether it is indeed the (single) standard model Higgs boson or if it is accompanied by new physics \cite{Higgs3}. These precision studies of Higgs boson properties will require
lepton collisions on the TeV energy scale in a future linear accelerator.
As compared with protons, leptons are favorable in this regard due to
their point-like structure. Proposals for corresponding linear $e^+e^-$ \cite{elcollprop} or
$\mu^+\mu^-$ \cite{mucolldesign} colliders have been put forward, but not been approved yet due to the considerable financial demands. However, the large variety of interesting particle reactions in high-energy lepton collisions renders their realization desirable.\\
In view of the ever increasing dimensions of high-energy particle
colliders, alternative acceleration concepts are nowadays being developed.
They might help to keep future facilities within reasonable size limits by
complementing existing technologies. Among them, laser acceleration
belongs to the most promising candidates \cite{Malka}. Super-intense laser pulses
available today provide acceleration gradients up to $\sim 1$\,GeV/cm,
exceeding the value at  conventional particle accelerators by three orders
of magnitude. Via relativistic laser-plasma interaction, electron beams with energies in GeV order have been generated along a 3 cm long acceleration capillary
\cite{Leemans}. A meter-scale plasma-wakefield accelerator has been able to double the energy of a 42 GeV electron beam \cite{Blumenfeld}. Proposals for relatively compact future laser-driven TeV
lepton colliders have been presented, accordingly \cite{Leemans2}.\\
Also, direct laser-particle acceleration for application in high-energy
physics is under active scrutiny. In particular, while moving inside an
intense laser field, particles can be accelerated temporarily to enormous
energies which may be exploited if the collisional event occurs in the
presence of the field \cite{McDonald,KarenEPL,Eminov}. Experimental implementations of high-energy QED processes via the interaction of matter with intense lasers have been realized \cite{Burke, Cowan, Schwoerer, Chen, kaminskireview, dipiazzareview}. Furthermore, theorists are currently developing new schemes for the employment of next-generation lasers in the probing of fundamental physics beyond the standard model \cite{Gies}.\\
Against this background, we study in this Letter electroweak processes in high-energy lepton collisions arising from a staged scheme
combining conventional with direct laser acceleration \cite{weakfootnote}. The incident
particles are assumed to be pre-accelerated by conventional means to a
center-of-mass (c.m.) energy which is high but still below the required
reaction threshold. The acceleration gradient provided by a collinear
superimposed laser field further enhances the c.m. energy, rendering possible the
particle reaction of interest. 
As a specific example, we consider the process {$\ell^+\ell^-\to HZ^0$} in
the presence of a super-intense laser wave (cf. Fig.~\ref{Scheme}). A detailed calculation within
the framework of laser-dressed quantum field theory is presented and its partially unexpected outcome as well as the
technical requirements for an experimental realization are discussed. Throughout this Letter, we use a natural units system with $\hbar=c=4\pi\varepsilon_0=1$.
\\

We assume a linearly polarized, monochromatic plane laser wave with four-potential $A_l(\phi):=a_l\cos\phi$, 
with $a_l=(0,a,0,0)$, and $\phi=(kx)$ being the four-product of the laser wave vector $k=(\omega, 0,0,\omega)$ and the space-time coordinate $x$. 
The transition amplitude for the process {$\ell^+\ell^-\to HZ^0$} inside the laser field is given by 
\begin{align}\label{TransitionAmplitude}
 \mathscr S&=\frac{-ig}{2\cos\theta_W}\iiint   \overline\psi_{+}(x) \gamma_\mu(g_V-g_A\gamma_5)\psi_{-}(x)\nonumber\\
&\times \frac{4\pi\e^{iq(x-y)}}{(2\pi)^4}\,\,\,\frac{-ig^{\mu\nu}+\frac{iq^\mu q^\nu }{M_Z^2}}{q^2-M_Z^2}\nonumber\\
&\times \frac{igM_Zg_{\nu\rho}}{\cos\theta_W} \sqrt{\frac{1}{4E_ZE_H}}\epsilon^*_\rho(P_Z)\e^{i(P_Z+P_H)y}
d^4xd^4y d^4q\,,
\end{align}
where the first line contains the leptonic current $J_\mu$, the second line represents the propagator of the virtual $Z^0$ boson \cite{propagatorfootnote}, and the third line describes the outgoing particles. Here, $g=e/\sin\theta_W$ is the weak coupling constant ($\theta_W$ being the Weinberg angle), $g_V$ and $g_A$ are the leptonic weak neutral current coupling constants, $g_{\mu\nu}$ is the metric tensor, $\gamma_\mu$ are the Dirac matrices, and $\gamma_5=i\gamma_0\gamma_1\gamma_2\gamma_3$. Within the Furry picture, the influence of the laser field on the charged leptons is accounted for by employing dressed states $\psi_\pm$ which are exact solutions to the Dirac equation in the presence of the plane electromagnetic wave $A_l(\phi)$. Carrying no electric charge, the created $H$ and $Z^0$ bosons are not influenced by the laser field and thus are described by free states, i.e. the polarization state $\epsilon(P_Z)$ for the $Z^0$ boson and 1 for the (scalar) Higgs particle. Their masses and four-momenta are $M_Z,P_Z=(E_Z,\vec P_Z)$ 
and $M_H,P_H=(E_H,\vec P_H)$, respectively. \\
Employing Feynman slash notation, the dressed leptons in Eq.~\eqref{TransitionAmplitude} are described by the Volkov states \cite{Berestetskii} 
\begin{equation}\label{VolkovStatesElectrons}
   \psi_{\pm}(x)=\sqrt{\frac{1}{2q^0_\pm}}\left(1\pm\frac{e\dirac k\dirac A}{2(kp_\pm)}\right)u_{p_\pm s_\pm}
\e^{iS}
\end{equation}
with the elementary charge unit $e$, lepton mass $m$, free Dirac spinors $u_{p_\pm s_\pm}$ and the classical action
\begin{equation}\label{action}
 S=\pm(q_\pm x) + \frac{e(a_lp_\pm)}{(kp_\pm)}\sin\phi \pm \frac{e^2a^2}{8(kp_\pm)}\sin2\phi\,.
\end{equation}
Note that Eqs. \eqref{VolkovStatesElectrons} and \eqref{action} contain laser-dressed four-momenta
\begin{equation}\label{dressemomenta}
 q^\mu_\pm = p^\mu_\pm + \xi^2\frac{m^2}{2(kp_\pm)}k^\mu
\end{equation}
which, depending on the dimensionless laser intensity parameter $\xi=ea/\sqrt{2}m$, may be considerably larger than the free momenta $p_\pm$. Note that the (quantum mechanical) laser-dressed momenta $q_\pm$ can be obtained from the classical particle momenta in the field by averaging over one laser cycle, see e.g. Section III in \cite{dipiazzareview}.\\
The leptonic current $J_\mu$ can be treated by standard QED procedure. Using the abbreviation $\Gamma_\mu:=\gamma_\mu(g_V-g_A\gamma_5)$, it is proportional to 
\begin{align}\label{leptcurrent}
 J_\mu\propto 
&\quad \bar u_{p_+s_+}\left(\Gamma_\mu + \frac{e}{2}\left(\frac{\dirac a_l\dirac k\Gamma_\mu}{(kp_+)}-\frac{\Gamma_\mu\dirac k\dirac a_l}{(kp_-)}\right)\cos\phi\right.\nonumber\\
&\qquad\qquad \left. - \frac{e^2\dirac a_l \dirac k \Gamma_\mu \dirac k\dirac a_l}{4(kp_+)(kp_-)}\cos^2\phi\right)u_{p_-s_-}\nonumber\\
&\times\text{e}^{-i\left(\beta_1\sin\phi+\beta_2\sin(2\phi)\right)}\text{e}^{-i(q_++q_-)x}
\end{align}
 with 
\begin{equation}\label{BessArg}
 \beta_1=\Bigl(\frac{e(a_lp_+)}{(kp_+)}-\frac{e(a_lp_-)}{(kp_-)}\Bigr),
\beta_2=\frac{e^2a^2}{8}\Bigl(\frac{1}{(kp_+)}+\frac{1}{(kp_-)}\Bigr)\,.
\end{equation}
The periodic functions $f(\phi):=\text{e}^{-i\left(\beta_1\sin\phi+\beta_2\sin(2\phi)\right)}$, $\cos\phi f(\phi)$ and $\cos^2\phi f(\phi)$ in Eq.~\eqref{BessArg} can be expanded in Fourier series of the form $\sum_n J_n(\beta_1, \beta_2)\text{e}^{-in\phi}$ where the coefficients can be expressed by generalized Bessel functions $J_n(\beta_1,\beta_2)$ \cite{Reiss}. This expansion allows for integrating Eq.~\eqref{TransitionAmplitude} over $x$ which for each $n$ yields a four-momentum conserving $\delta-$function at the leptonic vertex, $\delta(q_++q_-+nk-q)$. The $y-$integral similarly yields $\delta(P_Z+P_H-q)$. Hence, after integration over $q$, the transition amplitude becomes
\begin{align}\label{transampli}
 \mathscr{S} =\sum_n &\frac{(2\pi)^5}{4\sqrt{q_+^0q_-^0E_ZE_H}}\frac{g^2M_Z}{\cos^2\theta_W}\mathcal M_\mu^n\frac{-ig^{\mu\nu}+\frac{iq_n^\mu q_n^\nu}{M_Z^2}}{q_n^2-M_Z^2}\nonumber\\
&\times g_{\nu\rho}\epsilon^*_\rho(P_Z)\delta(q_n-P_H-P_Z)
\end{align}
with the virtual $Z^0$ boson's four-momentum $q_n:=q_++q_-+nk$ and a complex spinor-matrix product $\mathcal M_\mu^n$ containing the expansion of the leptonic current. The remaining $\delta$-function in Eq.~\eqref{transampli} represents the conservation of four-momentum in the process. Note that it contains the laser-dressed, i.e. boosted four-momenta $q_\pm$ of the incoming leptons. The summation index $n$ denotes the number of laser photons absorbed or emitted (if $n<0$) at the leptons' vertex. \\

In order to obtain a cross section, the transition amplitude \eqref{transampli} has to be squared, averaged over the initial spins and summed over the polarization states of the outgoing $Z^0$ boson, divided by the incoming particle flux $\abs{\vec j}$ and a unit time $T$, and integrated over the final particles' momenta:
\begin{align}\label{DifferentialCrossSection}
 d^6\sigma=\frac{1}{4}\sum_{s_+}\sum_{s_-}\sum_{\text{pol.}}\frac{\left|\mathscr{S}\right|^2}{T\abs{\vec j}}\frac{d^3\vec P_Z}{(2\pi)^3}\frac{d^3\vec P_H}{(2\pi)^3}\,.
\end{align}
Note that the particle flux $\abs{\vec j}$ \cite{Berestetskii} contains the laser-dressed momenta $q_\pm$. The spin sum in Eq.~\eqref{DifferentialCrossSection} can be evaluated by standard trace technology. 
Due to the $\delta-$function in Eq.~\eqref{transampli}, the square of the transition amplitude $\mathscr S=\sum_n\mathscr S_n$ can be written as $\left|\mathscr S\right|^2=\sum_n\left|\mathscr S_n\right|^2$ and the total cross section decomposes into a sum over partial cross sections $\sigma_n$, 
\begin{equation}\label{sigsum}
 \sigma=\sum_n\sigma_n\,.
\end{equation}
Note that, because $q_\pm\to p_\pm$ for $a_l\to0$, Eq.~\eqref{sigsum} reproduces the field-free cross section of the process in the limit of vanishing laser field.\\
 
The influence of the laser field on the collision energy is two-fold: Firstly, the laser-dressed lepton momenta $q_\pm$ appear in the four-momentum conservation law of the process. For a particle which counterpropagates with the laser field, the laser-dressed energy amounts to $q^0\approx p^0$, while for a co-propagating particle one finds $q^0\approx (1+\xi^2)p^0$, which can be significantly larger than the incident energy $p^0$ (cf. Eq.~\eqref{dressemomenta} and \cite{McDonald, Chan}). An intuitive reason for this enormous energy gain is that, in its rest frame, the co-moving particle experiences a strongly enhanced acceleration time.

Secondly, the absorption of $n$ laser photons leads to additional four-momentum $nk$ in the lepton vertex. At high laser intensities, the number $n$ of absorbed laser photons can be very large, thus further increasing the available collision
energy. If the laser beam propagates along the collision axis (such that $\beta_1=0$), photon numbers as high as
$n_{\text{max}}\approx2\beta_2=p^0\xi^2/\omega$ give contributions to the cross section. Here, $p_+^0=p_-^0\equiv p^0$
has been assumed. The distribution of partial cross sections $\sigma_n$ even peaks at $n\approx n_{\text{max}}$ due to the properties of the
Bessel functions \cite{ftnoteBessel}. The collision energy in the presence of the laser field, $E_{\text{cm}}(n)=[(q_++q_-+nk)^2]^{1/2}$, is therefore not fixed
but varies with $n$, reaching a maximum value of 
\begin{eqnarray}\label{Evonnmax}
E_{\text{cm}}(n_{\text{max}}) = 2p^0\sqrt{1+2\xi^2}
\end{eqnarray}
which may substantially exceed the field-free collision energy $2p^0$. Therefore, superimposing a strong laser field may render the process {$\ell^+\ell^-\to HZ^0$} possible even if the incident free lepton energies lie below the reaction threshold \cite{ftnote_eq10}. \\

\begin{figure}
\includegraphics[width=0.4\textwidth]{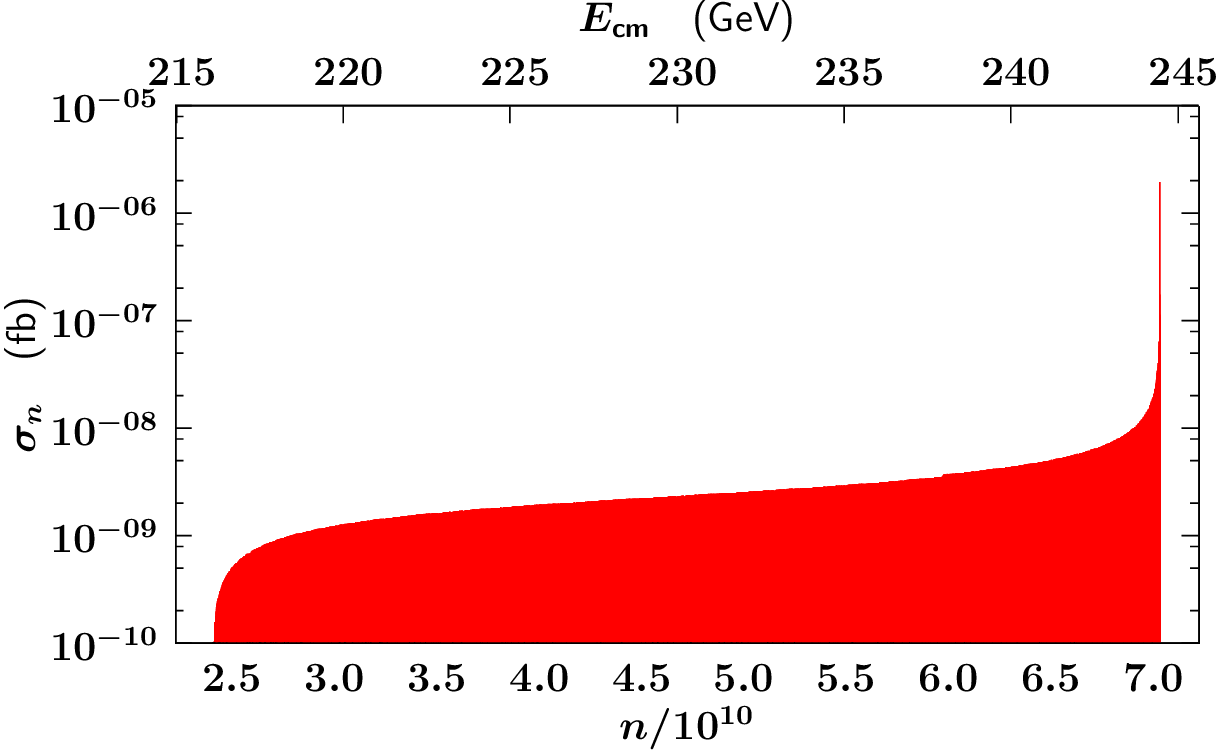}
 \caption{(Color online) Partial cross section $\sigma_n$ as function of the number of absorbed photons $n$ (lower $x$-axis)
 on a logarithmic scale. On the upper $x$-axis, the corresponding collision energy $E_{\text{cm}}(n)$ is shown. Free lepton energy $p^0_\pm\approx70$\,GeV, $\xi=1$, $\omega=1$\,eV. Note that only every $10^{6}$-th photon order is depicted. }\label{sigma_n_w1}
\end{figure}

As an example, Fig.~\ref{sigma_n_w1} shows the partial cross sections $\sigma_n$ for the process $\mu^+\mu^-\to H Z^0$ in the presence of a super-intense
near-optical laser field ($\sim10^{22}$\,W/cm$^2$). The Higgs boson mass is taken as 125 GeV \cite{Higgs}. Without the laser field, the total cross section attains a maximum of $\approx$212 fb at a c.m. energy of about {245 GeV} \cite{Donoghue}. Here, we chose the incident muon energy $p^0$ in such a way that the energy gain from the laser field boosts the total collision energy for the maximum number of absorbed photons \eqref{Evonnmax} to this value, $E_{\text{cm}}(n_{\text{max}})\approx245$\,GeV. This is the case for $p^0\approx70$\,GeV. Note that there occurs a cut-off towards the left side in Fig.~\ref{sigma_n_w1} where the number of absorbed photons does not suffice to exceed the reaction threshold. Summation over all contributing partial cross sections yields $\sigma\approx38$\,fb. Note that this cross section is an average over a wide range of collision energies. Therefore, it does not compare directly to a field-free cross section at a well-defined energy. However, the rate of Higgs creation events 
per unit time $\mathcal R = \sigma\abs{\vec j}$, with the particle flux $\abs{\vec j}\approx 2/V[1+m^2\xi^4/4p_0^2(1+\xi^2)]$ and the interaction volume $V$, can be accessed experimentally in the usual way \cite{Higgs, Burke}.\\

In view of an experimental realization of the described scheme certain conditions
regarding the collision parameters should be met \cite{McDonald}. Firstly, the quasi-momentum
in Eq.~\eqref{dressemomenta} strictly holds for an infinitely extended laser wave and is meaningful in a finite laser pulse only if the particles experience at least
one full oscillation cycle. This is certainly guaranteed for the counterpropagating
antimuon in our setup but for the copropagating muon it is required that the laser
pulse length exceeds $\Delta z=2\gamma^2(1+\xi^2)\lambda$ \cite{Chan}, with $\gamma=p^0/m$ and the laser wavelength $\lambda$. In addition,
the transversal extent $\Delta x$ of the laser beam along the polarization direction should cover the complete muon trajectory, i.e.
$\Delta x\gtrsim\xi\gamma\lambda$. Note that focusing and trapping of the particles in the field is a common issue in laser acceleration schemes. It can be achieved by tailored pulse profiles and auxiliary fields \cite{leptonbeam}.\\ 
Secondly, the quasi-momentum \eqref{dressemomenta} is maximal for perfect
alignment of the lepton and laser beams. This can be realized only if the
angular spread in the incident particle beam is low, $\Delta\theta\ll1/\gamma$. If
$\Delta\theta>\xi/\gamma$, only a small fraction of the muon beam
is efficiently accelerated. Typical envisaged beam divergences for muon colliders are in the order of 1 mrad \cite{bking2}. 
\\
Finally, the motion of the leptons can be affected by radiation damping due to Compton scattering in the field which, in principle, limits the applicable laser intensities. However, this effect is suppressed by the large muon mass. The relevant parameter measuring the impact of radiation damping for the muon and antimuon, respectively, is given by $R_\pm=\alpha\xi^2(kp_\pm)/m^2$ \cite{ADP}, with the QED fine structure constant $\alpha$. For the parameters of Fig. \ref{sigma_n_w1} we obtain $R_-\ll R_+\approx10^{-7}$, rendering radiation damping effects of minor importance. 
\\Due to the small total cross section which is typical in weak interaction processes, there will be a number of background processes, like e.g. $\mu^+\mu^-\rightarrow e^+e^-$, which would occur in any collision experiment. The laser field, however, allows for some additional channels. In particular, $e^+e^-$ pairs can be created from the collision of the counterpropagating antimuons with the laser photons via the nonlinear Bethe-Heitler effect \cite{ Mueller}. This field-induced background process estimates to be rather weak, leading to only one created pair per antimuon during the interaction time.\\

In Tab.~\ref{tabelle}, exemplary laser beam parameters are given for different collision constellations, where the lepton energies are such that the absorption of $n_{\text{max}}$ photons leads to the collision energy with the maximum field-free cross section. The required laser pulse power and pulse energy are calculated for elliptically-shaped laser pulses \cite{Yariv} where
the extension parallel to the electric field is assumed to
cover the whole lepton trajectory and perpendicular to
it, the radius is assumed to be $\lambda$ \cite{divergence}. Besides the parameters for the process $\mu^+\mu^-\rightarrow HZ^0$ considered in Fig. \ref{sigma_n_w1}, we also list the parameters for $\xi=0.5$. Here the required laser pulse power and energy but also the boost effect on the leptons are smaller. For comparison, laser-boosted $e^+e^-$ collisions are listed as well. For the latter, high values of $\xi$ are much easier to obtain since the required laser intensities scale as $\sim m^2$. However, larger Lorentz factors than for muons, $\sim1/m$, are necessary, leading to vastly increased spatiotemporal beam dimensions and thus huge beam energies. Moreover, Compton scattering and radiation losses impose more stringent limitations. Therefore, inducing electroweak reactions by laser-boosting of $\mu^+\mu^-$ collisions appears more promising with regard to experimental implementation.\\
As can be seen from the pulse durations given in Tab. 1, the field contains many laser cycles ($\mathcal{O}(10^6)$ for the parameters from column 1). This justifies the assumption of a monochromatic laser field of infinite duration underlying our calculations. We note that, for much shorter pulses, it would be required to take into account the shape of the finite laser pulse which leads to quantitative modifications \cite{pulsecitations}.
\\

\begin{table}
\centering
 \begin{tabular}{rcccc}
\hline\hline
 & $\mu^+\mu^-$ & $\mu^+\mu^-$ &  $e^+e^-$ & $e^+e^-$  \\ \hline 
Intensity param. $\xi$ & $1$ & $0.5$ & $1$ & $0.5$  \\ 
Intensity (W/cm$^2$) & $7.6\!\times\! 10^{22}$ & $1.9\!\times\! 10^{22}$ &    $1.8\!\times\! 10^{18}$ & $4.4\!\times\! 10^{17}$ \\ 
Optimum $p^0$ (GeV) & $70$ & $100$ & $70$ & $100$ \\ 
Lorentz factor $\gamma$ & $670$ & $945$ & $1.4\!\times\!10^5$  & $2.0\!\times\!10^5$   \\
Extension$ || \vec{E}$ (mm) & $0.8$ & $0.6$ & $170$ & $120$  \\
Pulse duration (ns)& $7$ & $9$ & $3\!\times\!10^5$ & $4\!\times\!10^5$  \\ 
Pulse power (PW)& $2.3\!\times\!10^3$ & $400$ & $11$ & $2$  \\ 
Pulse energy (GJ)& $16$ & $3.6$ & $3.4\!\times\!10^3$ & $750$  \\ \hline\hline
 \end{tabular}
\caption{Required laser parameters for the considered example {$\mu^+\mu^-\to HZ^0$} for the laser wavelength $\lambda\approx1.2$ $\mu$m. The incident lepton energy $p^0$ is such that Eq. \eqref{Evonnmax} yields $E_{\text{cm}}\approx245$~GeV, where the field-free cross section is maximal. A laser beam with an elliptical cross sectional area is assumed, with the extensions parallel to the electric and magnetic field components given by $\Delta x = \xi\gamma\lambda$ and $\Delta y = \lambda$, respectively. The pulse duration is $2\gamma^2(1+\xi^2)\lambda/c$. Power and energy of the pulse are derived from its spatiotemporal extensions. For comparison, a smaller laser intensity parameter as well as the required parameters for $e^+e^-$ collisions are listed. }\label{tabelle}
\end{table}

Let us put the laser parameters given in the first two columns of Tab. \ref{tabelle} into perspective. With modern petawatt laser systems, peak intensities in the order of $10^{22}$\,W/cm$^2$ have indeed been achieved \cite{yanovsky}. The focal extensions ($\sim\mu$m) and pulse durations ($\sim\text{fs}$), however, lie substantially below the required level. Further development towards 10 PW machines is currently being pushed forward \cite{Vulkan, Apollo}, with the goal of reaching a few 100 PW within the future Extreme Light Infrastructure \cite{ELI}. Even exawatt laser systems 
are already being considered \cite{Mourou}. Present high-power lasers are able to generate pulses with energies in the MJ range and durations of about 10\,ns \cite{NIF}. Thus, while the laser parameters required for an efficient boosting of $\mu^+\mu^-$ collisions are certainly very challenging, the ongoing progress in high-power laser technology offers prospects to reach them.\\

While the laser field may strongly enhance the c.m. energy in laser-boosted lepton colliders,
it reduces the collider's luminosity: the interaction area is increased in the polarization direction and
the number of reacting particles is decreased due to the small focus in the perpendicular direction. Assuming an initial bunch population of
$2\times10^{12}$ muons, a bunch radius of 26 $\mu$m \cite{Alsharoa}, and a laser field
with the parameters of the first column in Tab. \ref{tabelle}, we can estimate that the large
extension of the field in $x$-direction leads to a luminosity reduction by a factor
of $26\,\mu{\rm m}/\Delta x \approx 3 \times 10^{-2}$. In addition, due to the small extension in $y$-direction,
only a fraction of the initial muons is sufficiently accelerated which decreases the
luminosity by another factor of $\Delta y/26\,\mu{\rm m}\approx 4 \times 10^{-2}$. Regarding the collision frequency,
one may assume that the repetition rate of the laser can reach the repetition rate $\sim 15$ Hz
of the envisioned muon collider \cite{Alsharoa} and that the lepton and laser beams are perfectly
synchronized. Thus, all in all,  we obtain a luminosity loss due to the laser field of approximately
three orders of magnitude.\\

\begin{figure}
\includegraphics[width=0.4\textwidth]{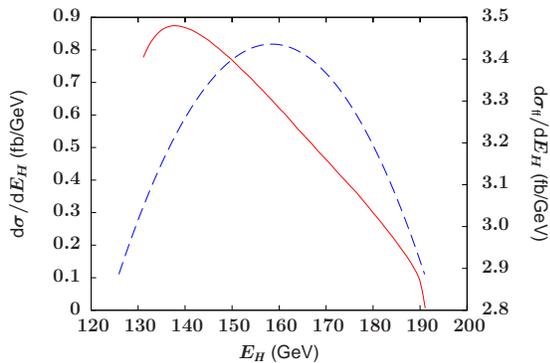}
 \caption{(Color online) Differential cross section as function of the produced Higgs boson's energy for $\xi=1$, $p^0\approx70$ GeV, and $\omega=10$ keV (red solid line, left $y$-axis). For comparison, the energy spectrum for a field-free collision with $p_+^0\approx70$ GeV, $p_-^0=q_-^0+n_{\text{max}}\omega\approx212$ GeV is depicted (blue, dashed line, right $y$-axis). }\label{dsigma_dEH}
\end{figure}

We note that higher laser frequencies can be beneficial for our scheme. Assuming that ultraviolet laser systems with sufficient intensity and power become available, the required laser pulse energy in Tab. \ref{tabelle} could be reduced by an order of magnitude by utilizing a laser beam with $\omega = 10$\,eV (keeping $\xi=1$) since the said pulse energy $\sim a^2\omega^2\Delta x\Delta y\Delta z$ scales linearly with the laser wavelength. Alternatively, application of this frequency along with a pulse extent of $\Delta y = 10\lambda$ would lead to a luminosity loss of only two orders of magnitude (while keeping the laser pulse energy unchanged).\\

Apart from the boost of the leptons' collision energy, the presence of the laser field may lead to several additional interesting effects. The broad distribution in Fig. \ref{sigma_n_w1} and supporting calculations for other parameters give evidence that the Higgs boson energy spectrum is asymmetric, with a tendency to smaller energies as compared to the corresponding field-free case. This is illustrated by a calculation with $\xi=1$, $p^0\approx70$ GeV and $\omega=10$ keV for which the energy distribution is shown in Fig.~\ref{dsigma_dEH}. Note that the total cross section and behavior of the spectrum are practically independent of the photon energy provided that $\omega\ll m$. Compared to a field-free cross section with comparably asymmetric lepton energies, smaller Higgs boson energies are favored. This is because fewer partial cross sections of photon order $n$ yield sufficiently high energy to contribute for larger Higgs boson energy. For the same reason, the spectrum of the Higgs boson's emission angle is asymmetric and tilted towards the propagation direction of the laser. 
Besides that, the laser field will also affect the electrically charged decay products of the Higgs boson, e.g. by channeling them into narrow angular regions \cite{erik}, thus possibly rendering their detection easier. We also note that, in comparison to the field-free case, the total cross section obtained in the laser-boosted reaction $\mu^+\mu^-\to HZ^0$ may be somewhat less sensitive to variations of the Higgs boson mass which is connected to the broad range of collision energies due to the many photon numbers that are involved (cf. Fig. \ref{sigma_n_w1}). \\
In conclusion, we have shown that the upgrading of the existing and the development of new laser facilities -- in fruitful conjunction with the well-established conventional acceleration techniques -- might bring new opportunities in the physics of elementary particles such as to Higgs boson creation. As a proof-of-principle experiment, the laser-boosted QED process $e^+e^-\rightarrow\mu^+\mu^-$ could be implemented with present-day technology, utilizing intense nanosecond laser pulses with an energy of kJ order.\\

\begin{acknowledgments}
 Useful input by Karl-Tasso Kn\"opfle, Karen Z. Hatsagortsyan, Omri Har-Shemesh, Matthias Ruf, and Felix Mackenroth is gratefully acknowledged.
\end{acknowledgments}


\begin{thebibliography}{1}
\bibitem{propagatorfootnote} Note that, due to the large collision energy required in the process, the finite $Z^0$ boson width can be neglected in the propagator.
\bibitem{ftnote_eq10}  The variation of the collision energy with the photon order $n$ corresponds, in classical terms, to the temporal oscillations of the lepton energies in the field. Accordingly, one can show that \eqref{Evonnmax} can also be derived from a consideration based on the classical dynamics of the particles in the field.
\bibitem{divergence} In order for the laser pulse to maintain this shape over the whole required extension $\Delta z$, very advanced focusing techniques are necessary in order to counteract the beam divergence. Waveguides consisting of hollow capillaries \cite{wgcapp} or plasma waveguides \cite{wgcite} might be promising candidates for obtaining this goal.
\end{thebibliography}
\end{document}